\documentclass[conference]{IEEEtran}
\IEEEoverridecommandlockouts
\usepackage{cite}
\usepackage{soul}
\usepackage{graphicx}
\usepackage{textcomp}
\usepackage{amsmath}
\usepackage{amssymb}
\usepackage{amsfonts}
\usepackage{textcomp}
\usepackage{multirow}
\usepackage{tabularx}
\usepackage{caption}
\usepackage{float}
\usepackage{graphicx, picture, calc}
\usepackage{epstopdf}
\usepackage{caption}
\usepackage{mathtools}
\usepackage{booktabs}
\usepackage{enumerate}
\usepackage{footmisc}
\usepackage{algorithm}
\usepackage{algpseudocode}
\usepackage{multirow}
\usepackage{pifont}
\usepackage{color}
\usepackage{mathptmx}
\usepackage{txfonts}
\usepackage{tabularx}
\usepackage{latexsym}
\usepackage{stfloats}
\usepackage{footnote}
\usepackage{balance}
\usepackage{babel}
\usepackage{url}

\def\BibTeX{{\rm B\kern-.05em{\sc i\kern-.025em b}\kern-.08em
    T\kern-.1667em\lower.7ex\hbox{E}\kern-.125emX}}
\begin{document}

\title{Secrecy Performance Analysis of Space-to-Ground Optical Satellite Communications}
\author{\IEEEauthorblockN{Thang V. Nguyen\IEEEauthorrefmark{1},
Thanh V. Pham\IEEEauthorrefmark{3}, 
Anh T. Pham\IEEEauthorrefmark{4},
and Dang T. Ngoc\IEEEauthorrefmark{1}}
\IEEEauthorblockA{\IEEEauthorrefmark{1}Wireless Systems and Applications Lab.,
Posts and Telecommunications Institute of Technology,
Hanoi, Vietnam}
\IEEEauthorblockA{\IEEEauthorrefmark{3}Department of Mathematical and Systems Engineering, Shizuoka University, Japan}
\IEEEauthorblockA{\IEEEauthorrefmark{4}Department of Computer Science and Engineering, The University of Aizu, Japan}
}

\maketitle

\begin{abstract}
To cope with the scarcity of radio frequency (RF) counterparts, free-space optics (FSO) based on satellite communications technologies have recently gained a lot of interest. Firstly, this investigates the security of space-to-ground intensity modulation/direct detection FSO satellite communications in various effects from the physical world, such as propagation loss, beam misalignment, cloud attenuation, and fading caused by atmospheric turbulence. Next, A wiretap channel composed of a genuine broadcaster Alice (i.e., the satellite), a legitimate user Bob, and an eavesdropper Eve is considered over turbulent channels characterized by the Fisher-Snedecor $\mathcal{F}$ distribution. In addition, the closed-form expressions of the secrecy performance metrics, including average secrecy capacity, secrecy outage probability, and strictly positive secrecy capacity, are derived in our study. Finally, the numerical findings indicate that the satellite altitude of 600 km should be used to achieve a secrecy outage of roughly 50\% for the whole turbulence regime. In addition, high levels of cloud liquid water content (CLWC) could improve the secrecy performance as it is shown that, the average secrecy capacity increases by about 12\% when the CLWC increases from 1 mg$/$m$^3$ to  2 mg$/$m$^3$. 

\end{abstract}


\section{Introduction}
Over the past decade, satellite communications (SatCom) have emerged as the new frontier beyond fifth-generation (B5G) networks, which extend service coverages and transmit high-data rates to places where telecommunications infrastructure is lacking or unavailable, such as rural areas, remote areas, and even in the ocean \cite{Kodheli2020}. Recent advancements in space communications have been fueled by the popularity of low Earth orbit (LEO) satellite mega-constellations \cite{Illi_OCJS2020, Thang_TAES2023}. 
\begin{figure}[htbp]
\centerline{\includegraphics[width = 0.7\linewidth]{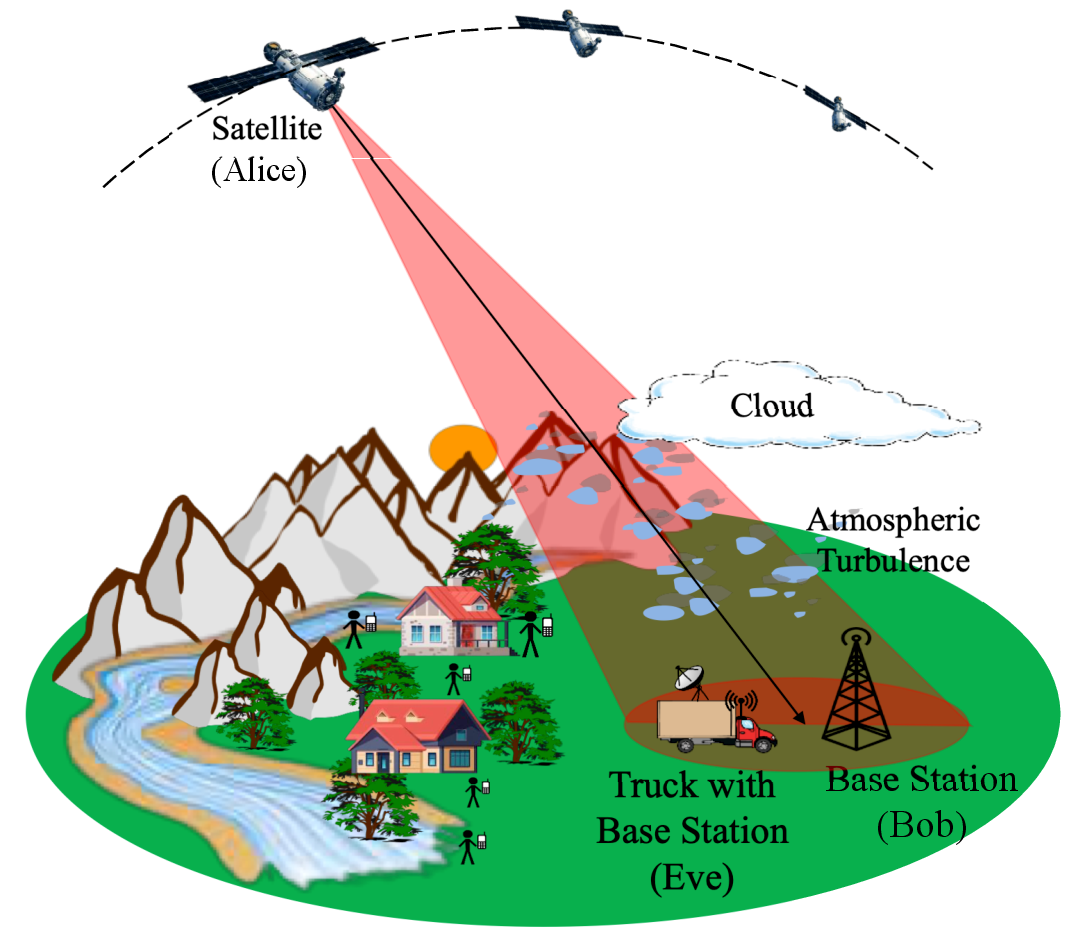}}
\caption{An example of a wiretap FSO-based satellite communications link.}
\label{fig}
\end{figure}
Currently, RF bands between the L-band (1-2 GHz) and the Ka-band (26-40 GHz) are being used extensively for SatCom. However, the increasing demand for higher data-rate transmission has necessitated the use of optical frequencies. In addition to the increased bandwidth, free-space optics (FSO)-based SatCom also offers better energy efficiency, lightweight transceiver design, and license-free spectrum \cite{Kaushal_Survey2017, Toyoshima2021}. As a result, it is expected that the use of FSO and SatCom together provides a possible remedy for future communication networks.

While the design, performance analysis, and feasibility of FSO-based SatCom systems have been intensively investigated in the literature, there has been little research on privacy and security. In the case of terrestrial FSO, due to the relatively short link ranges (i.e., several hundred meters to a few kilometers), the beam footprint at the receiver 
is sufficiently small, making it difficult for eavesdropping by malicious users. In the case of SatCom, however, the transmission distances are in the order of at least a few hundred kilometers. Hence, the radius of the beam footprint at the ground station can be up to several hundred meters. This poses a considerable security threat as an eavesdropper (Eve) can be in the beam footprint without being detected by the legitimate user (Bob). 

Traditional security measures are performed at the upper layers of the OSI model by means of key-based cryptography. Aside from this, physical layer (PHY) security has been gaining a lot of interest as a novel approach to prevent eavesdropping without using sophisticated data encryption techniques \cite{Saber_WCL2017}. In this regard, a few works in the literature have studied the secrecy performance of FSO-based SatCom \cite{Illi_OCJS2020, Ai_PJ2019, Ma_TCOM2022}. In \cite{Illi_OCJS2020}, the PHY secrecy performance was investigated for high throughput hybrid terrestrial-satellite system with FSO feeder links. The intercept probability metric was used to evaluate the system performance over the gamma-gamma turbulence channel. In \cite{Ai_PJ2019}, the authors used the gamma-gamma model to formulate the secrecy outage performance of an FSO satellite cooperative system. In a different attack setting, a group of aerial eavesdroppers tries to eavesdrop on satellite-unmanned aerial vehicles (UAVs) communications \cite{Ma_TCOM2022}.  Then, the authors investigated the probability of secrecy outage performance of the uplink transmission. 

Different from these previous works, an all-optical satellite-to-ground wiretap channel is examined in this study. As shown in Fig. \ref{fig}, we consider a wiretap channel consisting of a legitimate satellite (Alice), a legitimate base station (Bob), and an eavesdropper (Eve) (e.g., a truck mounted with a base station, which can move within the beam footprint area). Then, new closed-form expressions for the average secrecy capacity (ASC), the secrecy outage probability (SOP), and the strictly positive secrecy capacity (SPSC) are derived, taking into account the atmospheric loss, misalignment, cloud attenuation, and atmospheric turbulence-induced fading. Especially in this study,  atmospheric turbulence fading channels modeled by the Fisher-Snedecor $\mathcal{F}$ distribution are considered. As verified in \cite{Badarneh_CL2021}, the $\mathcal{F}$ distribution exhibits a better fit to the experimental data than the commonly used gamma-gamma and log-normal distributions over all turbulence conditions. Extensive simulations are performed to illustrate the impact of various system and environmental parameters on the system's secrecy performance. 



\section{Channel Model} \label{channel_model}
An optical signal propagating from the satellite to the ground station experiences various impairments due to the atmospheric and clouds. As a result, the end-to-end channel model can be expressed as 
\begin{align}
h = h_{\text{a}}h_\text{s} h_\text{c} h_\text{t},
\end{align}
where $h_\text{a}$ is the deterministic atmospheric loss, $h_{\text{s}}$ is the misalignment, $h_\text{c}$ is the cloud attenuation, and $h_\text{t}$ is the turbulence-induced fading. 

\subsection{Atmospheric Loss}
Due to the absorption of laser beam energy and alteration of the light's direction by air atoms and aerosol particles, the optical signal weakens as it traverses from the satellite to Bob and Eve. 
The well-known Beer-Lambert law is used to calculate the atmospheric path loss, which is a function of transmission distance and is given by
\begin{align}
h_a = \exp \left( { - {\sigma_{tr}}L} \right)  \exp \left( { - {\sigma_{st}}H_{st}\sec(\xi)} \right),
\end{align}
where $L=(H_s - H_g)/\cos{(\xi)}$ is the propagation length of the FSO link in which $\xi$ is the satellite's zenith angle, $H_s$ is the satellite altitude, and $H_g$ is the high of the ground station (GS). $\sigma_{tr}$ and $\sigma_{st}$ are the attenuation coefficient of the troposphere layer (about 20 km above Earth's surface) and the stratosphere layer (from 20 to 48 km above Earth's surface). $H_{st}$ is the vertical extent of the stratosphere layer. Table \ref{attenuation} shows the values of $\sigma_{tr}$ and $\sigma_{st}$ at the optical wavelength of $\lambda = 1550$ nm \cite{Fidler_JSAC2010}.

\begin{table}[hthp]
\centering
\caption{Attenuation coefficient for two atmospheric layers \cite{Giggenbach_ICSCE, Thang_JOC}.}
\label{attenuation}
\begin{tabular}{|c|c|l|l|}
\hline
\multirow{8}{*}{\rotatebox{90}{Attenuation coefficient}} & \multirow{4}{*}{\begin{tabular}[c]{@{}l@{}}Troposphere layer\\ $\sigma_{tr}$ (Km$^{-1}$)\end{tabular}}  & Heavy fog           & 28.75       \\ \cline{3-4} 
 &  & Light fog           & 4.6         \\ \cline{3-4}   &   & Haze                & 0.966       \\ \cline{3-4}&     & Clear air           & 0.0989      \\ \cline{2-4}  & \multirow{4}{*}{\begin{tabular}[c]{@{}l@{}}Stratosphere layer\\ $\sigma_{st}$ (Km$^{-1}$)\end{tabular}} & Extreme volcanic    & $10^{-1}$   \\ \cline{3-4} & & High volcanic       & $2\times10^{-2}$ \\ \cline{3-4} & & Moderate volcanic   & $8\times10^{-4}$ \\ \cline{3-4} & & Background volcanic & $10^{-4}$   \\ \hline
\end{tabular}
\end{table}
\subsection{Misalignment}
To calculate the impact of beam misalignment at the receiver, a Gaussian beam profile and a circular detecting aperture are assumed. The normalized spatial distribution of the optical intensity  at a distance, $L$, can be expressed by
\begin{align}
    {I_{beam}}\left( {{\rho _e};L} \right) = \frac{2}{{\pi \omega _L^2}}\exp \left( { - \frac{{2\left\| {{\rho _e}} \right\|}}{{\omega _L^2}}} \right),
\end{align}
where $\omega_L=\omega_0\sqrt{1+\epsilon\left(\frac{\lambda L}{\pi \omega_0^2} \right)^2}$ is the beam size at the distance $L$, $\omega_0 = (2\lambda)/(\pi \theta)$ is the beam waist at $L=0$ with $\theta$ being the divergence angle, and $\epsilon$ is given in \cite{Thang_TAES2023}. $\rho_e$ is the radial vector from the center of the beam footprint, and $\left\| {}\cdot \right\|$ defines the expression of Euclidean norm. The beam spreading loss is quantified by the fraction of power collected by the detector $h_{s}(\cdot)$. It depends on not only the beam size but also the relative position between the center of the detector and the beam footprint, which is known as a pointing error. Denoting $r$ as the pointing error, $h_{s}(\cdot)$ can be determined as
\begin{align}
    {h_{s}}\left( {r;L} \right) = \int\limits_A {{I_{beam}}\left( {\rho  - r;L} \right)d\rho } ,
\end{align}
where $A$ is the area of the detecting aperture. The Gaussian form of $h_{s}(\cdot)$ is written as
\begin{align}
    {h_{s}}\left( {r;L} \right) \approx {A_0}\exp \left( { - \frac{{2{r^2}}}{{\omega _{Leq}^2}}} \right)\exp \left( { - \frac{{2{d_E^2}}}{{\omega _{Leq}^2}}} \right),
\label{misalignment}
\end{align}
where $\omega _{Leq}^2 = \omega _L^2(\sqrt \pi  erf(v))/2v\exp ( - {v^2})$ defines the equivalent beam width at the destination. ${A_0} = {[erf(v)]^2}$ and $v = (\sqrt \pi  D_r)/(2\sqrt 2 {\omega _L})$ in which $D_r$ is the diameter of the detection aperture at the GS, $r$ is the distance between the center of the beam footprint and the detector. $A_0$ denotes the fraction of collected power at $r=0$. Moreover, $d_E$ is the distance between Eve's and Bob's positions on the receiver plane.

\subsection{Cloud Attenuation}
According to recent investigations, clouds have a severe negative impact on the FSO link \cite{Hoang2021, Pham2023}. System performance and availability might suffer significantly in the presence of cloud covering due to high cloud liquid water content (CLWC). The Beer-Lambert law can be used to express cloud attenuation as
\begin{align}
{h_\text{c}} = \exp \left( { - {\alpha _2}{d_\text{c}}} \right),
\end{align}
where $d_c$ is the propagation slant path and $\alpha_2$ is the attenuation coefficient and can be expressed as 
\begin{align}
{\alpha _2} = \frac{{3.91}}{{V\left[ \text{km} \right]}}{\left( {\frac{{\lambda \left[ \text{nm} \right]}}{{550}}} \right)^{ - q}},
\end{align}
where $q$ is the coefficient of Kim’s model \cite{Hoang2021}.
The visibility can be determined based on the cloud droplet number concentration $N_\text{c}$ and CLWC $M_\text{c}$ as followed
\begin{align}
V = \frac{{1.002}}{{{{\left( {{N_\text{c}} \times {M_\text{c}}} \right)}^{0.6473}}}} .
\end{align}
The typical values of $N_\text{c}$ and $M_\text{c}$ for multiple types of clouds are presented in \cite{Erdogan2021}.
\subsection{Atmospheric Turbulence-induced Fading}
Atmospheric turbulence-induced fading is a consequence of changes in the refractive index of the air
due to the inhomogeneity of temperature and pressure in the atmosphere. It causes fluctuations in the received optical intensity, thus significantly reducing the performance of the free-space channel. To measure the turbulence strength, $\sigma_R^2$ is the Rytov variance which defines weak, moderate, and strong turbulence corresponding to $\sigma_R^2<1$, $\sigma_R^2 \approx 1$, and $\sigma_R^2>1$. In the case of plane wave propagation, the Rytov variance, denoted by $\sigma_R^2$, is given as \cite{Gaussian_beam_2005}  and can be shown as
\begin{align}
\sigma _R^2 = 2.25{k^{7/6}}{\sec ^{11/6}}\left( \xi  \right)\int\limits_{{H_g}}^{{H_s}} {C_n^2\left( h_t \right){{\left( {h_t - {H_g}} \right)}^{5/6}}dh_t},
\label{Rytov}
\end{align}
where $C_n^2 (h_t)$ is the variation of the refractive index structure parameter described by the Hufnagel-Valley model \cite[Eq.~(19)]{Thang_PJ_2021} and can be expressed as
\begin{align}
    C_n^2&(h_t) =  0.00594\left(\frac{v_{\text{wind}}}{27} \right)^2 \left(10^{-5}h_t \right)^{10} \exp\left(-\frac{h_t}{1000}\right) \nonumber \\
    &+ 2.7 \times 10^{-16} \exp\left(-\frac{h_t}{1500}\right) + C_n^2(0) \exp\left(-\frac{h_t}{100}\right),
\end{align}
where $C_n^2(0)$ is the ground level turbulence, and $v_{\text{wind}}$ (m/s) is the root mean squared wind speed. 

In the literature, several statistical distributions, for example, the log-normal distribution, gamma-gamma distribution, and Malaga distribution, have been examined to model turbulence-induced fading with different degrees of fitness. In this work, the fading modeled by the Fisher-Snedecor $\mathcal{F}$ distribution is employed due to its excellent agreement with the measurement data. 
The probability distribution function (PDF) of the turbulence-induced fading coefficient $h_t$ can be expressed as \cite{Peppas_JLT2020} 
\begin{align}
    f_{h_t}(h_t)=\frac{a^a (b-1)^b h_t^{a-1}}{\mathcal{B}(a,b)(a h_t + b - 1)^{a+b}},
\label{f_h_t}
\end{align}
where $\mathcal{B} (\cdot)$ is the beta function \cite{Gradshteyn2007}. The parameters $a$ and $b$ of the $\mathcal{F}$-distribution is given by \cite{Peppas_JLT2020}  
\begin{align}
a = \frac{1}{{\exp \left( {\sigma _{\ln \mathcal{S}}^2} \right) - 1}}, 
\end{align}
and
\begin{align}
b = \frac{1}{{\exp \left( {\sigma _{\ln \mathcal{L}}^2} \right) - 1}} + 2,
\end{align} 
where ${\sigma _{\ln \mathcal{S}}^2}$ and ${\sigma _{\ln \mathcal{L}}^2}$ are the corresponding small-scale and large-scale log-irradiance variances can be found in \cite{Gaussian_beam_2005}. 
In the following, \eqref{f_h_t} can be rewritten by supporting of \cite[Eq.~(8.4.2.5)]{Prudnikov1986} 
\begin{align}
    {f_{{h_t}}}\left( {{h_t}} \right) \!=\! \frac{{{a^a}{h_t^{a - 1}}}}{{\mathcal{B}\left( {a,b} \right){{\left( {b - 1} \right)}^a}\Gamma \left( {a + b} \right)}}G_{1,1}^{1,1}\left( {\frac{a}{{b - 1}}h_t\left| {\begin{array}{*{20}{c}}{1 - a - b}\\0\end{array}} \right.}\!\!\!\! \right),
\end{align}
where $G^{m, n}_{p, q}(\cdot)$ is the Meijer-G function \cite{Gradshteyn2007}.  
To obtain the cumulative distribution function (CDF) of the $h_t$, \cite[(26)]{Adamchik1990} and \cite[Eq.~(9.31.5)]{Gradshteyn2007} are used together with some algebraic manipulations; the closed-form of CDF can be written as 
\begin{align}
    {F_{{h_t}}}\left( {{h_t}} \right) = \frac{1}{{\mathcal{B}\left( {a,b} \right)\Gamma \left( {a + b} \right)}}G_{2,2}^{1,2}\left( {\frac{a}{{b - 1}}h_t\left| {\begin{array}{*{20}{c}}
{1 - b}&1\\a&0\end{array}} \right.} \right).
\label{F_h_t}
\end{align}
\section{Secrecy Performance Analysis} \label{performance}
In this section, the FSO-based satellite system performance is investigated over the presented composite channel. The average secrecy capacity and strictly positive secrecy capacity are two key secure metrics derived from a closed-form expression.
\subsection{System Model}
We assume a communication link employing the on-off keying (OOK) modulation in the FSO-based satellite communications, in which the electrical signal received at the receiver $U$\footnote{The subscript `$U$' is used to denote Bob and Eves as users in general. When necessary, the subscript `$B$' is used to refer to Bob while `$E$' is used to refer to Eve.} can be expressed as 
\begin{align}
    r_U = h_U x + n_U,
\end{align}
where $h_U$ is the channel coefficient considered at $U$, $x \in \{0, 2P_t\}$ is the transmitted power intensity taken as a symbol drawn equiprobably from an OOK constellation,  $P_t$ is the average transmitted power, and $n_U$ is the corresponding signal-dependent additive white Gaussian noise (AWGN) with variances 
$\sigma^2_{n,U}$ \cite{Phuc_TCOM2020}. Hence, the received electrical signal-to-noise ratio (SNR) over a fading channel can be  defined as 
\begin{align}
    {\gamma_U} = \frac{{2P_t^2h_U^2}}{{\sigma _{n,U}^2}}=4\overline{\gamma}_Uh_U^2,
\end{align}
where $\overline{\gamma}_R$ is the electrical SNRs in the case of no fading. By transforming the random variable h, the PDF and CDF of $\gamma_U$ can be given as
\begin{align}
    f_{\gamma_U}(\gamma_U) = \frac{1}{2\mathcal{B}(a,b)\Gamma(a+b)} \gamma_U^{-1} G_{1,1}^{1,1}\left[ {\left.\frac{a}{b-1} \sqrt{\frac{\gamma_U}{4\overline{\gamma}_U}} \right|\begin{array}{*{20}{c}}{1 - b}\\a\end{array}} \right],
\label{f_gamma}
\end{align}
and 
\begin{align}
    F_{\gamma_U}(\gamma_U) = \frac{1}{\mathcal{B}(a,b)\Gamma(a+b)} G_{2,2}^{1,2}\left[ {\left.\frac{a}{b-1} \sqrt{\frac{\gamma_U}{4\overline{\gamma}_U}} \right|\begin{array}{*{20}{c}}{1 - b}&1\\a&0\end{array}} \right],
\label{F_gamma}
\end{align}
respectively. 
\subsection{Secrecy Performance Metrics}
\subsubsection{Average Secrecy Capacity (ASC)}
In AWGN channels, the instantaneous secrecy capacity is given as the difference between the capacities of Bob's and Eve's channels as
\begin{align}
    C_S(\gamma_B, \gamma_E) = \left[\text{log}_2(1+\gamma_B) - \text{log}_2(1+\gamma_E)\right]^+, 
\end{align}
where $[x]^+ \overset{\Delta}{=} \text{max}(x, 0)$, $\text{log}_2(1+\gamma_B)$ and $\text{log}_2(1+\gamma_E)$ correspond to the capacity of Bob and Eve channels, respectively. Over the fading channel, the average secrecy capacity (ASC) can be defined as
\begin{align}
    \text{ASC}=\int_0^\infty \int_0^\infty C_S(\gamma_B,\gamma_E)f_\gamma(\gamma_B,\gamma_E) ~ d\gamma_B d\gamma_E.
    \label{ASC_exact}
\end{align}
Due to the independency between $\gamma_B$ and $\gamma_E$, the joint PDF of $\gamma_B$ and $\gamma_E$ can be shown as $f(\gamma_B,\gamma_E)=f_{\gamma_B}(\gamma_B)f_{\gamma_E}(\gamma_E)$. The ASC in \eqref{ASC_exact} can be rewritten as
\begin{align}
    \text{ASC} &= \frac{1}{\text{ln2}} \left(\int_0^\infty \text{ln}(1+\gamma_B)f_{\gamma_B}(\gamma_B)\int_0^{\gamma_B} f_{\gamma_E}(\gamma_E)~d\gamma_E d\gamma_B \right. \nonumber \\ 
    & ~~~~~~\left. - \int_0^\infty \text{ln}(1+\gamma_E)f_{\gamma_E}(\gamma_E)\int_{\gamma_E}^\infty f_{\gamma_B}(\gamma_B)~d\gamma_B d\gamma_E\right) \nonumber \\
    & = \frac{1}{\ln 2}\left(\underbrace{\int_0^\infty \text{ln}(1+\gamma_B)f_{\gamma_B}(\gamma_B) F_{\gamma_E}(\gamma_B)~d\gamma_B}_{A_1} \right. \nonumber \\ 
    & ~~~~~~\left. + \underbrace{\int_0^\infty \text{ln}(1+\gamma_E)f_{\gamma_E}(\gamma_E) f_{\gamma_B}(\gamma_E)~d\gamma_E}_{A_2} \right. \nonumber \\
    & ~~~~~~\left. - \underbrace{\int_0^\infty \text{ln}(1+\gamma_E)f_{\gamma_E}(\gamma_E)~d\gamma_E}_{A_3} \right).  
\end{align}

Using \eqref{f_gamma}, \eqref{F_gamma}, and the identity $\text{ln}(1+\gamma)=G_{2,2}^{1,2}\left[ {\gamma \left| {\begin{array}{*{20}{c}}1&1\\1&0\end{array}} \right.} \right]$ \cite[(8.4.6.5)]{Prudnikov1986}, $A_1$ is given by
\begin{align}
&{A_1} = \frac{1}{{2{{\left[ {\mathcal{B}\left( {a,b} \right)\Gamma \left( {a + b} \right)} \right]}^2}}}\int\limits_0^\infty  {{\gamma_B ^{ - 1}}G_{2,2}^{1,2}\left[ {\gamma_B \left| {\begin{array}{*{20}{c}}1&1\\1&0\end{array}}\!\!\right.} \right]} \nonumber\\
&\times G_{1,1}^{1,1}\left[ {\frac{a}{{b + 1}}\sqrt {\frac{\gamma_B }{{4\overline{\gamma}_B  }}} \left| {\begin{array}{*{20}{c}}{1 - b}\\a
\end{array}} \right.}\!\!\right] G_{2,2}^{1,2}\left[ {\frac{a}{{b + 1}}\sqrt {\frac{\gamma_B }{{4\overline{\gamma}_E}}} \left|\!\!{\begin{array}{*{20}{c}}{1 - b}&1\\a&0
\end{array}} \right.}\!\!\!\right]d\gamma_B.
\label{A1_integral}
\end{align}
By applying \cite[(07.34.21.0081.01)]{Wolfram}, the integral in \eqref{A1_integral} can be expressed in terms of the extended generalized bivariate Meijer G-function (EGBMGF), which is shown in \eqref{A1_closed} on top of the next page. Similarly, an expression for $A_2$ is given in \eqref{A2_closed}.
\begin{figure*}
\begin{align}
{A_1} = \frac{{{{\left( {{2^{a + b}}} \right)}^2}}}{{16{\pi ^2}{{\left[ {\mathcal{B}\left( {a,b} \right)\Gamma \left( {a + b} \right)} \right]}^2}}} G_{2,2:2,2:4,4}^{2,1:2,2:2,4}\left[ {\begin{array}{*{20}{c}}0&1\\0&0\end{array}\left| {\begin{array}{*{20}{c}}
{\frac{{1 - b}}{2}}&{\frac{{2 - b}}{2}}\\{\frac{a}{2}}&{\frac{{a + 1}}{2}}
\end{array}\left| {\left. {\begin{array}{*{20}{c}}
{\frac{{1 - b}}{2}}&{\frac{{2 - b}}{2}}&{\frac{1}{2}}&1\\
{\frac{a}{2}}&{\frac{{a + 1}}{2}}&0&{\frac{1}{2}}
\end{array}} \right|\frac{{{a^2}}}{{4\overline{\gamma}_B  {{\left( {b - 1} \right)}^2}}},\frac{{{a^2}}}{{4\overline{\gamma}_E  {{\left( {b - 1} \right)}^2}}}} \right.} \right.} \right].
\label{A1_closed}
\end{align}
\begin{align}
    {A_2} = \frac{{{{\left( {{2^{a + b}}} \right)}^2}}}{{16{\pi ^2}{{\left[ {\mathcal{B}\left( {a,b} \right)\Gamma \left( {a + b} \right)} \right]}^2}}}G_{2,2:2,2:2,2}^{2,1:2,2:2,2}\left[ {\begin{array}{*{20}{c}}1&2\\1&1\end{array}\left| {\begin{array}{*{20}{c}}
{\frac{{1 - b}}{2}}&{\frac{{2 - b}}{2}}\\{\frac{a}{2}}&{\frac{{a + 1}}{2}}
\end{array}\left| {\left. {\begin{array}{*{20}{c}}{\frac{{1 - b}}{2}}&{\frac{{2 - b}}{2}}\\
{\frac{a}{2}}&{\frac{{a + 1}}{2}}\end{array}} \right|\frac{{{a^2}}}{{4\overline{\gamma}_B  {{\left( {b - 1} \right)}^2}}},\frac{{{a^2}}}{{4\overline{\gamma}_E  {{\left( {b - 1} \right)}^2}}}} \right.} \right.} \right].
\label{A2_closed}
\end{align}
\line(1,0){520}
\end{figure*}

\begin{table}[t]
\caption{System Parameters.}
\begin{center}
\begin{tabular}{|l|c|c|}
\hline
\textbf{Parameters}&{\textbf{Symbol}} &{\textbf{Value}} \\
\cline{2-3} 
\hline
Optical wavelength & $\lambda$&  $1550$ nm  \\
The high of ground users & $H_g$&  $10$ m  \\
Attenuation coefficient of troposphere layer & $\sigma_{tr}$ & 0.002 dB/km \\
Attenuation coefficient of stratosphere layer & $\sigma_{st}$ & 0.001 dB/km \\
Cloud liquid water content & $M_c$ & 1 mg/m$^3$ \\
The considered length of cloud  & $d_c$ & 2 km \\
Receiver aperture diameter & $D_r$ & 5 cm \\
Wind speed & $v_{\text{wind}}$ & 21 m/s \\
Target secrecy rate & $C_T$ & 0.5 bits/s/Hz \\
\hline
\end{tabular}
\label{Sys_para}
\end{center}
\end{table}
\begin{figure}[t]
\centering
\includegraphics[height = 0.63\linewidth, width = .9\linewidth]{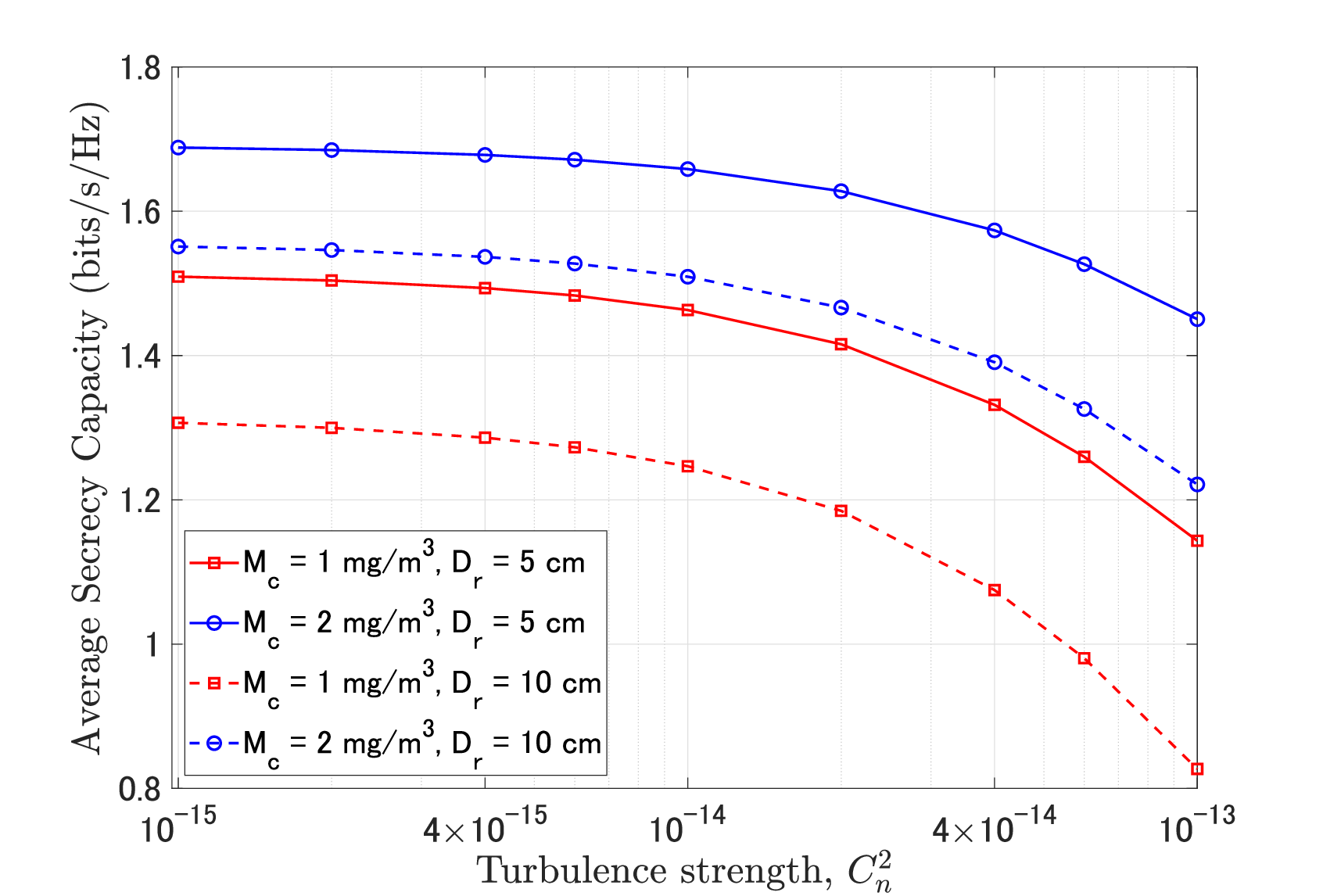}
\caption{ASC versus turbulence strength with $\xi=60^\circ$ and $H_s=800$ km.}
\label{ASC_tur}
\end{figure}

Finally, using \cite[Eq. (21)]{Adamchik1990} and after some mathematical manipulations, the closed-form expression for $A_3$ can be shown as
\begin{align}
   {A_3} = \frac{{{2^{a + b}}}}{{4\pi B\left( {a,b} \right)\Gamma \left( {a + b} \right)}}G_{4,4}^{4,3}\left[ {\frac{{{a^2}}}{{4{{\left( {b - 1} \right)}^2}\overline{\gamma}_E  }}\left| {\begin{array}{*{20}{c}}{\frac{{1 - b}}{2}}&{\frac{{2 - b}}{2}}&0&1\\
{\frac{a}{2}}&{\frac{{a + 1}}{2}}&0&0\end{array}} \right.} \right].
\label{A3_closed}
\end{align}
Using \eqref{A1_closed}, \eqref{A2_closed}, and \eqref{A3_closed}, the closed-form expression for the ASC is obtained.
\subsubsection{Secrecy Outage Probability (SOP)} 
A secrecy outage event is defined when the instantaneous secrecy capacity falls below a required threshold. Denote $C_T > 0$ as the outage threshold; the SOP is then expressed by 
\begin{align}
    \text{SOP} &=\text{Pr}\left(C_s(\gamma_B, \gamma_E)<C_T\right) \nonumber \\ 
 &= \text{Pr} \left[ \text{log}_2\left(\frac{1+4\overline{\gamma}_B h_B^2}{1+4\overline{\gamma}_E h_E^2} \right) < C_T\right] \nonumber \\
    & = \int\limits_0^\infty F_{h_B}\left(\sqrt{\frac{2^{C_T}(1+4\overline{\gamma}_E h_E^2)-1}{4\overline{\gamma}_B}} \right)f_{h_E}(h_E)dh_E.
\label{SOP_exact}
\end{align}
Since an exact closed-form expression for \eqref{SOP_exact} is difficult to obtain, we aim at deriving a lower bound for the SOP. Specifically, since $\frac{2^{C_T}(1+4\overline{\gamma}_E h_E^2)-1}{4\overline{\gamma}_B} > \frac{2^{C_T\overline{\gamma}_E}h^2_E}{\overline{\gamma}_B}$, a lower bound for \eqref{SOP_exact} can be given by
\begin{align}
   \text{SOP}_\text{LB} =& \int\limits_0^\infty F_{h_B} \left( 2^{\frac{C_T}{2}} \sqrt{\frac{\overline{\gamma}_E}{\overline{\gamma}_B}}h_E \right) f_{h_E}(h_E) dh_E \nonumber \\ 
   =& \frac{{{a^a}}}{{{{\left[ {\mathcal{B}\left( {a,b} \right)\Gamma \left( {a + b} \right)} \right]}^2}{{\left( {b - 1} \right)}^a}}} \nonumber \\ 
   &\times \int\limits_0^\infty  h_E^{a - 1}G_{2,2}^{1,2}\left( {\left. {\frac{a}{{b - 1}}{2^{\frac{{{C_T}}}{2}}}\sqrt {\frac{{{{\overline \gamma  }_E}}}{{{{\overline \gamma  }_B}}}} {h_E}} \right|\begin{array}{*{20}{c}}{1 - b}&1\\a&0\end{array}}\right) \nonumber \\ 
   &\times G_{1,1}^{1,1}\left( {\left. {\frac{a}{{b - 1}}{h_E}} \right|\begin{array}{*{20}{c}}{1 - a - b}\\0\end{array}} \right)d{h_E} .
\label{SOP_LB}
\end{align}
With the help of \cite[(2.24.1.1)]{Prudnikov1986}, and after some algebraic manipulations, a closed-form expression for \eqref{SOP_LB} can be derived as
\begin{align}
\text{SOP}_\text{LB} &= \frac{1}{{{{\left[ {\mathcal{B}\left( {a,b} \right)\Gamma \left( {a + b} \right)} \right]}^2}}}G_{3,3}^{2,3}\left( {\left. {{2^{\frac{{{C_T}}}{2}}}\sqrt {\frac{{{{\overline \gamma  }_E}}}{{{{\overline \gamma  }_B}}}} } \right|\!\!\begin{array}{*{20}{c}}{1 - b}&1&{1 - a}\\a&b&0\end{array}} \!\!\!\right).
\label{SOP_closed}
\end{align}

\subsubsection{Strictly Positive Secrecy Capacity (SPSC)}
To stress the existence of the secrecy capacity, the strictly positive secrecy capacity (SPSC) serves as another benchmark in secure communications. Mathematically speaking, the SPSC can be given by \cite{Bloch_TIT2008}
\begin{align}
    \text{SPSC}&=\text{Pr}\{C_S(\gamma_B, \gamma_E) > 0 \} \nonumber \\
    &= 1 - \text{SOP}|_{C_T=0}.
\label{SPSC_eq}
\end{align}
From \eqref{SOP_exact}, it is deduced that the lower bound of SOP becomes the exact form when $C_T=0$. Thus, the exact closed-form expression of SPSC can be attained by substituting \eqref{SOP_closed} into \eqref{SPSC_eq} and setting $C_T=0$.

\section{Results and Discussions} \label{results}
In this section, simulation results are provided to illustrate the impact of the system's parameters on the secrecy performance. Unless otherwise noted, the simulation parameters are given in Table~\ref{Sys_para}. 

First, we explore in Fig.~\ref{ASC_tur} the effect of the turbulence strength on the ASC with $\xi=60$ degree and $H_s=800$ km in four different settings of CLWC and the diameter of the photodetector. 
From the figure, we can see that the ASC  is almost unchanged in the weak turbulence regime (i.e., $C_n^2$ is between $10^{-15}$ to $4\times10^{-15}~\text{m}^{-2/3}$), gradually decreases in the moderate regime ($4\times10^{-15}\to$ $4\times10^{-14}~\text{m}^{-2/3}$), and then quickly drops in the strong regime ($4\times10^{-14}\to$ $10^{-13}~\text{m}^{-2/3}$). 
This illustrates the significant effect of turbulence-induced fading on the ASC performance. In addition, weather-related issues, such as CLWC, also considerably impact the ASC performance. For example, at $C^2_n = 10^{-14}~\text{m}^{-2/3}$ and $D_r = 5$ cm, the ASC decreases by about 12\% when the CLWC reduces from 2 $\text{mg/cm}^3$ to 1 $\text{mg/cm}^3$.  
\begin{figure}[t]
\centering
\includegraphics[height = 0.63\linewidth, width = .9\linewidth]{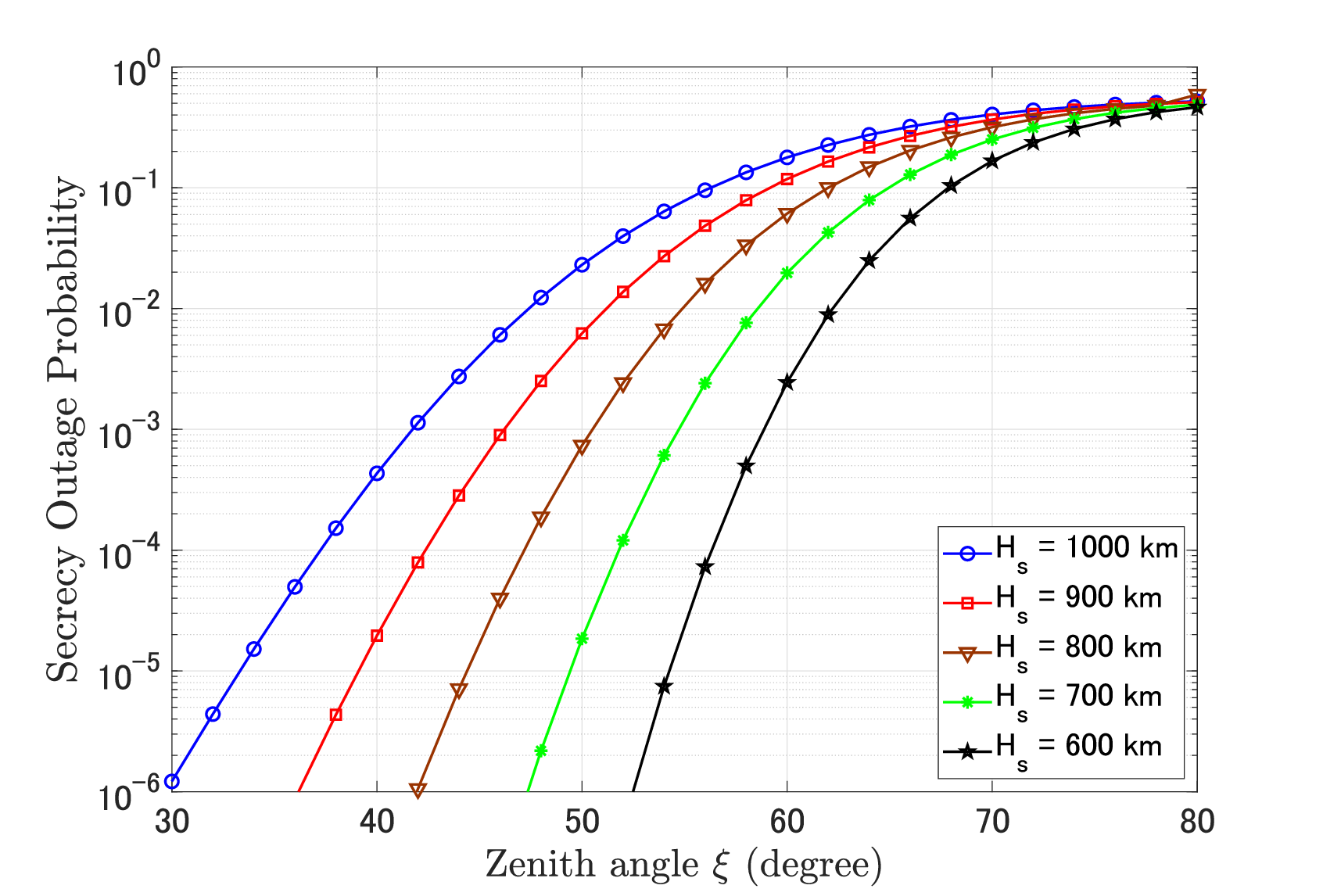}
\caption{SOP versus zenith angle with $d_E=2$ m, and $C_n^2=10^{-13}$ m$^{-2/3}$.}
\label{SOP_vs_Zenith}
\end{figure}

In Fig. \ref{SOP_vs_Zenith}, the SOP is shown as a function of the zenith angle $\xi$ for different values of the satellite altitude $H_s$ given the distance between Bob and Eve, $d_E=2$ m, and the turbulence strength, $C_n^2=10^{-13}$ m$^{-2/3}$. As it is expected, the lower orbit (i.e., smaller $H_s$) is, the lower the SOP becomes. Besides, as shown in \eqref{Rytov}, the zenith angle is a parameter directly related to turbulence strength. Thus, the SOP increases as the zenith angle rises due to the growth of the turbulence strength. For example, to achieve a target SOP of $10^{-3}$, the satellites at the altitude of 1000 km, 900 km, 800 km, 700 km, and 600 km should be at the zenith angle of $42^{\circ}$, $46^{\circ}$, $50^{\circ}$, $55^{\circ}$, and $59^{\circ}$, respectively.

\begin{figure}[t]
\centering
\includegraphics[height = 0.63\linewidth, width = .9\linewidth]{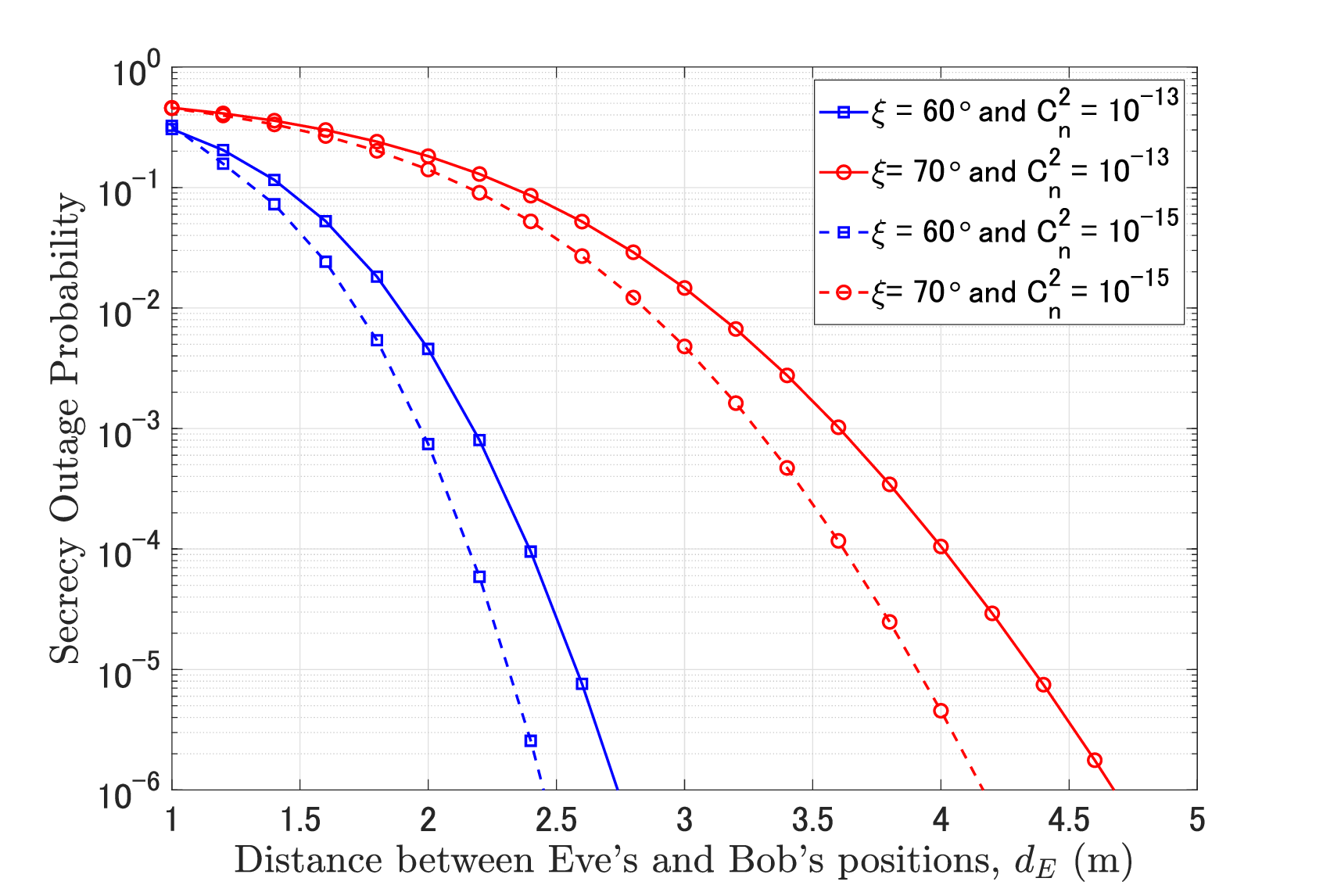}
\caption{The relationship between SOP and distance between Bob and Eve, $d_E$, with the satellite altitude of 600 km.}
\label{SOP_vs_distance}
\end{figure}

Figure \ref{SOP_vs_distance} shows the relationship between the SOP and the distance between Bob and Eve when the satellite altitude is 600 km. From this result, we would like to highlight three key observations. Firstly, the SOP decreases when Eve is far away from Bob in both strong turbulence, $C_n^2=10^{-13}~\text{m}^{-2/3}$, and moderate turbulence regimes, $C_n^2=10^{-15}~\text{m}^{-2/3}$. The reason is that Eve is more likely to benefit from a misalignment beam, especially at a further distance $d_E$, as shown in \eqref{misalignment}. Secondly, the stronger turbulence helps to reduce the SOP when Eve is away from Bob. Thirdly, the secrecy outage event is to occur when Eve is closer to Bob. For instance, the probability of secrecy outage is roughly $50\%$ for the whole turbulence conditions.   

\begin{figure}[t]
\centering
\includegraphics[height = 0.63\linewidth, width = .9\linewidth]{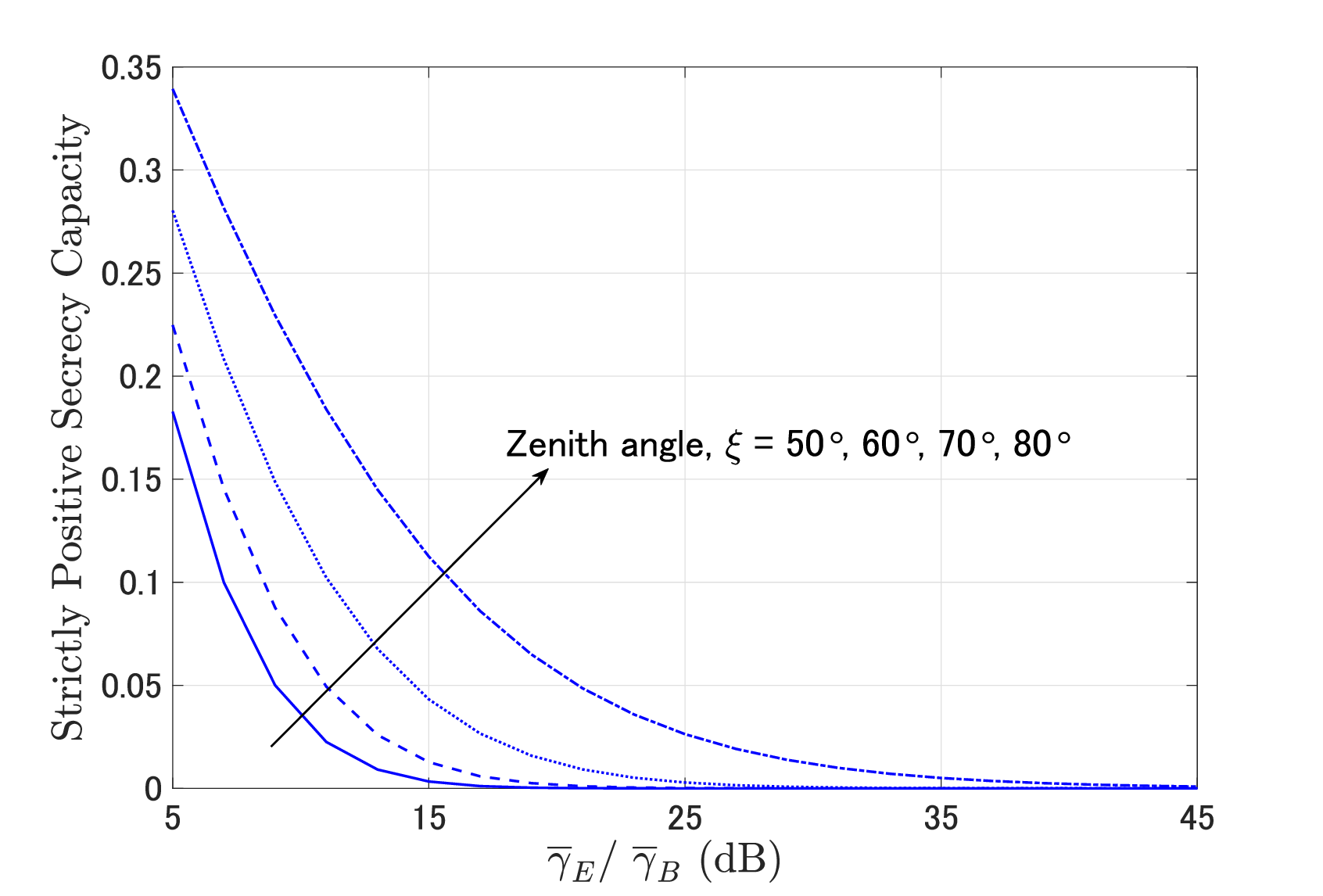}
\caption{SPSC versus average electrical SNR ratio with $\xi=50^{\circ}$, $H_s=600$ km, $d_E=4$ m, and $C_n^2=10^{-13}$ m$^{-2/3}$.}
\label{SPSC}
\end{figure}

Finally, we depict the SPSC with respect to the ratio between the average electrical SNRs of Eve and Bob, as shown in Fig.~\ref{SPSC}. Based on the derivation in \eqref{SPSC_eq}, we observed that the SPSC is enhanced by increasing the zenith angle, $\xi$, because the turbulence-induced fading strength increase when $\xi$ increases. In addition, it is shown that a lower SPSC corresponds to a higher SOP and vice versa. 

\section{Conclusions} \label{conclusion}
In this paper, we investigated the PHY security issues, which considered physical layer impairments such as attenuation from troposphere and stratosphere layers, the cloud effect, misalignment loss, and turbulence-induced fading, for FSO-based satellite communications. In particular, the atmospheric turbulence is modeled by Fisher-Snedecor $\mathcal{F}$ distribution, while misalignment loss considers the impact of distance between Eve's and Bob's positions on the secrecy performance of satellite communications. In addition, the closed-form expression for the secrecy outage probability and the strictly positive secrecy capacity are derived in this paper. Numerical results demonstrate the severe effect of turbulence strength and emphasize the impact of Eve's location on secrecy performance. 


\balance
\bibliographystyle{ieeetr}
\bibliography{Thang_lib}
\end{document}